# Lithographically-controlled liquid metal diffusion in graphene: Fabrication and magneto-transport signatures of superconductivity

S. Wundrack[1,2,3], M. Bothe[1,3], M. Jaime[1], K. Küster[5], M. Gruschwitz[4], Z. Mamiyev[4], P. Schädlich[4], B. Matta[5], S. Datta[5], M. Eckert[1,2], C. Tegenkamp[4], U. Starke[5], R. Stosch[1], H.W. Schumacher[1], T. Seyller[4], K. Pierz[1], T. Tschirner[1], A. Bakin[2,3]

[1] Physikalisch-Technische Bundesanstalt, Bundesallee 100, 38116 Braunschweig, Germany

[2] Institut für Halbleitertechnik, Technische Universität Braunschweig, Hans-Sommer Straße 66, D-38106 Braunschweig, Germany

[3] Laboratory of Emerging Nanometrology (LENA) der Technischen Universität Braunschweig, Langer Kamp 6 a/b, 38106 Braunschweig, Germany

[4] Institut für Physik, Technische Universität Chemnitz, Reichenhainer Str. 70, 09126 Chemnitz, Germany

[5] Max-Planck-Institut für Festkörperforschung, Heisenbergstraße 1, 70569 Stuttgart, Germany

**Abstract.** Metal intercalation in epitaxial graphene enables the emergence of proximity-induced superconductivity and modified quantum transport properties. However, systematic transport studies of intercalated graphene have been hindered by challenges in device fabrication, including processing-induced deintercalation and instability under standard lithographic techniques. Here, we introduce a lithographically controlled intercalation approach that enables the scalable fabrication of gallium-intercalated quasi-freestanding bilayer graphene (QFBLG) Hall bar devices. By integrating lithographic structuring with subsequent intercalation through dedicated intercalation channels, this method ensures precise control over metal incorporation while preserving device integrity. Magneto-transport measurements reveal superconductivity with a critical temperature $T_c^{onset} \approx 3.5$ K and the occurrence of a transverse resistance, including both symmetric and antisymmetric field components, which is attributed to the symmetric-in-field component to non-uniform currents. These results establish an advanced fabrication method for intercalated graphene devices, providing access to systematic investigations of confined 2D superconductivity and emergent electronic phases in van der Waals heterostructures.

## Introduction

The study of proximity effects in condensed matter physics has provided significant insights into how electronic properties can be tuned in materials. The proximity effect, particularly in superconducting systems, plays a crucial role in enabling and controlling of quantum phenomena in two-dimensional (2D) materials. Over the past decades, considerable efforts have been made to induce superconductivity in non-superconducting systems, including graphene, via various proximity effects[1].

Graphene, a single layer of carbon atoms arranged in a honeycomb lattice[2], has emerged as a model system for investigating quantum transport phenomena. While pristine graphene lacks an intrinsic superconducting gap due to its Dirac-like band structure, several approaches have been developed to induce superconductivity, including intercalation with metallic and superconducting elements[3], and the use of heterostructures with topological insulators[4]. Among these, intercalated epitaxial graphene has gained significant interest due to its tunable electronic properties and compatibility with large-scale device fabrication.

Cortés-del Río et al. have demonstrated the possibility of engineering of superconducting states in hydrogen-intercalated graphene, showing that precise intercalation control leads to localized superconducting regions with enhanced coherence lengths [5].

Additionally, Efetov & Einenkel reanalyzed in Ref [6], examined the role of electron-phonon interactions in intercalated graphene and suggested that such mechanisms could, under specific conditions, facilitate intrinsic superconductivity. While experimental verification remains pending, their theoretical predictions align with recent observations of enhanced superconducting coupling in gold-intercalated graphene [3].

In addition to intercalation, twisted-bilayer graphene (TBG) has emerged as a highly promising platform for investigating superconductivity. When two graphene layers are twisted at a magic angle (~1.1°), flat electronic bands form, leading to strong electronic correlations that give rise to superconductivity [7]. The discovery of superconductivity in TBG has provided a new avenue for understanding unconventional superconductors and has opened possibilities for engineering tunable superconducting states in graphene-based systems.

Moreover, recent work on gallium-intercalated graphene heterostructures has demonstrated its potential in proximity-induced superconductivity [4,8].

A major limitation in the study of intercalated graphene systems arises in the context of magneto-transport measurements. While techniques such as angle-resolved photoemission spectroscopy (ARPES) and scanning tunneling microscopy (STM) have provided crucial insights into the density of states (DOS) and electronic band structure, these methods do not directly probe charge transport properties. This has hindered efforts to establish a clear correlation between intercalation-induced band structure modifications and quantum transport phenomena, such as resistance characteristics, the suppression or modification of the quantum Hall effect (QHE), and the emergence of exotic quantum phases like the quantum spin Hall insulator (QSHI) or the anomalous quantum Hall effect (AQHE).

Beyond fundamental physics, precise transport characterization is essential for metrological applications of the quantum Hall effect (QHE), particularly as a quantum resistance standard [9]. The QHE plays a central role in the traceability of fundamental units such as ohm, farad, ampere, and kilogram in the revised International System of Units (SI)[10]. The ability to precisely control intercalation raises the question of whether new quantum effects, such as the quantum spin Hall insulator (QSHI) or the anomalous quantum Hall effect (AQHE), can emerge in metal-intercalated graphene under high magnetic fields. Understanding these effects is crucial for advancing 2D quantum transport physics and assessing the potential of intercalated graphene for next-generation quantum electrical standards.

To address this gap, our work presents a lithographical fabrication route of epitaxial graphene Hall bars, making intercalated graphene systems accessible for transport measurements. The proposed chip design preserves the integrity of the confined metal layers beneath graphene, enabling reliable magneto-transport experiments. Systematic magneto-transport measurements are carried out for investigations of superconductivity in gallium (Ga) intercalated quasi-freestanding bilayer graphene on 4H-SiC (SiC/2DGa/QFBLG), revealing a superconducting transition at $T_c^{onset} \approx 3.5$ K , the absence of QHE above $T_c^{onset}$ , and the occurrence of a transverse resistance, which exhibits both symmetric and antisymmetric field components attributed to the symmetric-in-field component to non-uniform currents. The ability to control the metal diffusion beneath graphene establishes a scalable platform for exploring proximity-induced superconductivity and other transport phenomena in metal intercalated epitaxial graphene.

## Methods

### A. Polymer-assisted graphene growth (PASG)

Monolayer epitaxial graphene was grown on a semi-insulating 4H-SiC(0001) substrates (10 × 5 mm$^2$) using the so-called polymer-assisted sublimation growth (PASG) method. The SiC wafer from II-VI Comp. has a nominal miscut of about 0.04°. The graphene samples were prepared according to the polymer-assisted sublimation growth (PASG) technique, which involves polymer adsorbates formed on the 4H-SiC surface by liquid phase deposition from a solution of a photoresist (AZ5214E) in isopropanol followed by sonication and short rinsing with isopropanol. The graphene layer growth was processed at 1750∘C (argon atmosphere ~1 bar, 6 min, zero argon flow) with pre-vacuum-annealing at 900∘C.

### B. Gallium-intercalation of epitaxial graphene using LiMIT

The commercial gallium (99.99% purity) was purchased from Heraeus. The Liquid Metal Intercalation Technique (LiMIT) was employed, involving the placement of a Ga droplet on the metal reservoir area of the graphene Hall bar device. The sample was gently heated to 70°C in a nitrogen atmosphere, while an external pressure was applied locally to the liquid metal droplet for 10 minutes.

### C. Confocal Raman spectroscopy

Confocal Raman spectroscopy was performed using a Witec Alpha 300 RA equipped with 300 grooves/mm grating, a Nd:YAG laser with an excitation wavelength of 488 nm (2.54 eV), and a 600 mm focal length. Raman mapping was conducted over a (20 × 20) μm² area with a step resolution of 0.2 μm. To prevent laser-induced damage or heating effects, the laser power was maintained below 2 mW.

### D. Angle-resolved photoelectron spectroscopy (ARPES)

ARPES measurements were performed with a NanoESCA (Scienta Omicron) using HeI radiation (21.2 eV). The samples were degassed around 150 °C prior to measurements.

### E. X-ray photoelectron spectroscopy (XPS)

XPS measurements were conducted using Al Kα radiation (h$\nu$ = 1486.6 eV) from a Specs XR50 X-ray source monochromatised with a Specs Focus 500 monochromator. All XPS measurements were taken at room temperature with a base pressure better than 3x10$^{-10}$ mbar and all the core level data were taken at $E_{\text{Pass}}$=10 eV.

### F. Scanning Electron Microscopy (SEM) and Four-point probe scanning tunneling microscopy (4pp-STM) measurements

SEM and 4pp-STM measurements were performed in UHV at a base pressure of 5 x $10^{-10}$ mbar. SEM images were measured at 15 kV, 1 nA. Without annealing, the investigated area was cleaned by removing adsorbates during continuous scanning with a STM tip. Transport measurements were performed by four electro-chemical etched tungsten tips. The probes were brought into the tunneling regime by feedback-controlled approaching. The ohmic contact was established by manually approaching each tip with feedback control switched off. For both, linear and square tip arrangements, a tip spacing of 400 nm was realized (700 nm in case of the first measurements) and a source current ranging from -1…+1µA applied.

### G. Magneto-transport measurement

The magneto-transport measurements were performed in a commercial Oxford Instruments bath cryostat with a 12 T superconducting magnet, an ADRET Electronique current source and a HP 3458 multimeter.

### H. Density functional theory (DFT)

The structural relaxations and electronic structure calculations were performed using the Quantum ESPRESSO package within the plane-wave pseudopotential approach. The relaxation and calculation of the Electron Localization Function (ELF) were performed within a 16 × 14 Å² supercell, consisting of a quasi-freestanding bilayer graphene (QFBLG) layer, a gallium monolayer (2DGa), and a single Ga adatom, forming the (SiC/2DGa monolayer/Ga adatom/QFBLG) system along the (0001) surface (see Fig. S5, SI). The SiC surface consists of two SiC layers, with the bottom layer passivated by hydrogen atoms along $(000\bar{1})$.

The lattice relaxation was carried out in multiple steps. Initially, during the relaxation of the (SiC/2DGa monolayer/QFBLG) system, all atoms below the Si-face were kept fixed, allowing only the Si atoms of the Si-face, the Ga monolayer, and the QFBLG layer to relax. The Ga atoms of the Ga monolayer were positioned in a 1×1 arrangement relative to the Si-face. The supercell was constructed and expanded based on the √3 × √3 reconstruction.

Following this relaxation step, a Ga adatom was placed in a hollow site position on the Ga monolayer, situated at the Ga monolayer/QFBLG interface, and the system was subsequently relaxed again.

The Perdew-Burke-Ernzerhof (PBE) parametrization of the generalized gradient approximation (GGA-PBE) was used for the exchange-correlation functional. Furthermore, projector augmented wave (PAW) pseudopotentials were employed, and the lattice relaxation was performed using a plane-wave energy cutoff of 816 eV, and the force convergence threshold of 0.003 eV/Å.

## Results & Discussion

## Design of liquid metal intercalation of epitaxial graphene Hall bars

Usually, the fabrication of metal intercalated graphene is done via lithography after the intercalation process as schematically shown in Fig. 1a-d. Due to the harsh conditions during reactive ion etching or chemical etching with potassium iodide or other etching chemicals, the intercalated metal layer is often removed (Fig. 1d) and a lot of defects are induced into the graphene (see also SI for detailed description of the standard lithography process, Fig. S1) leading to highly defective devices and bad transport properties. Therefore, we introduce an alternative approach in which a fully fabricated epitaxial graphene Hall bar device is intercalated with metal atoms in a subsequent step[11].

This method leverages the diffusion properties of the low-melting-point metal gallium enabling the intercalation of epitaxial graphene even at room temperature[12]. Figure 2a-e illustrates the key concept of the fabrication process.

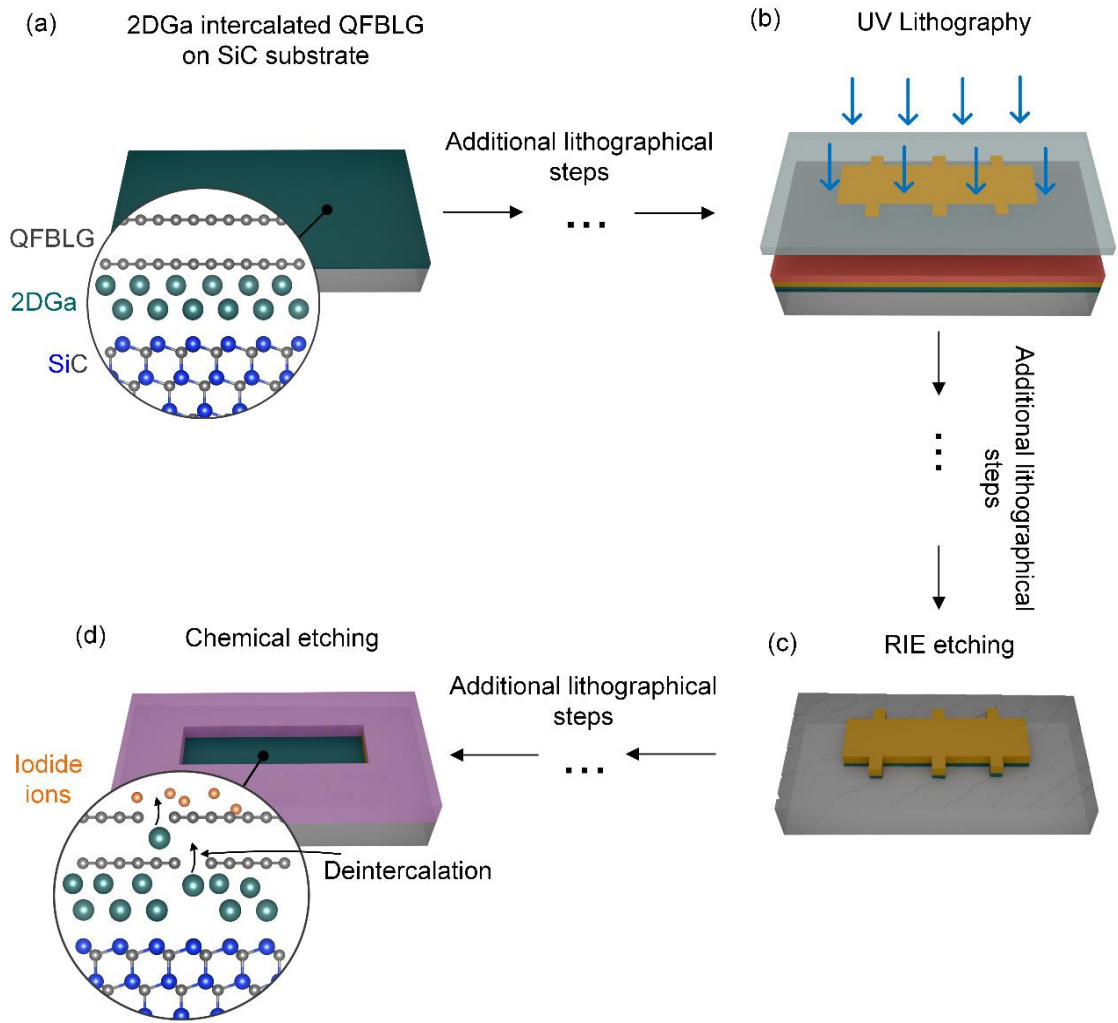

**Figure 1.** *Schematic illustration of bottleneck in lithographical fabrication of Ga-intercalated Hall bar devices. **a)** 2DGa-intercalated quasi-freestanding bilayer graphene (QFBLG) on a SiC substrate. **b)** UV lithography is performed to pattern the Hall bar geometry. **c)** Reactive ion etching (RIE) is applied to remove unwanted graphene, leaving the Hall bar intact. **d)** Chemical etching step leading to deintercalation of intercalated metals beneath QFBLG.*

The new fabrication approach of the graphene Hall bar design is based on the previously mentioned method using lithography, but starting with MLG instead of intercalated QFBLG (Fig. 1a). However, within the design structure, we incorporated predefined epitaxial graphene intercalation and metal reservoir areas that facilitate liquid metal diffusion (Fig. 2a, right inset). The intercalation channel serves as a connection between the metal reservoir and the graphene Hall bar device, facilitating efficient metal intercalation. The presence of these dedicated pathways ensures controlled and localized diffusion of the metal intercalant into the graphene Hall bar devices.

The graphene used in this study was synthesized via the polymer-assisted graphene growth (PASG) method[13], which is specifically designed to minimize undesirable effects such as step bunching of the SiC substrate and to improve the graphene quality by introducing polymer molecules as additional carbon sources within the epitaxial graphene growth. This optimization enables the formation of an atomically smooth epitaxial graphene layer on terraces with heights ranging from 0.50 to 0.75 nm (Fig. S2, see SI). By reducing anisotropy in the sheet resistance induced by substrate morphology and miscut angle, PASG enhances the electronic properties of epitaxial graphene[14].

Following graphene synthesis, device fabrication was carried out through lithographic structuring to define the Hall bar geometry and intercalation channels. Ensuring a controlled intercalation process required the precise structuring of a metal reservoir region, which serves as the source for gallium diffusion.

A critical step in the process involved the defect engineering of the metal reservoir region via controlled plasma treatment of the exposed epitaxial graphene (Fig. 2b), which facilitates subsequent intercalation. During this step, the reservoir area remains uncovered, while the intercalation channels and Hall bars are protected with a photoresist to prevent unintended defect formation. This selective protection is essential for preserving the intrinsic electronic properties of graphene in the Hall bar, as excessive defect introduction could degrade the performance[15], potentially impacting metal-intercalated graphene systems such as Ga-intercalated graphene and modifying quantum transport phenomena like the quantum Hall effect (QHE).

While high-temperature annealing can reduce the defect density, a residual defect concentration remains inherent, which can lead to metal atom deintercalation. Furthermore, the low intrinsic defect density of as-grown epitaxial graphene inhibits efficient gallium intercalation, necessitating additional defect introduction via plasma treatment. Details of the defect tuning employed in this method are provided in the supplementary material.

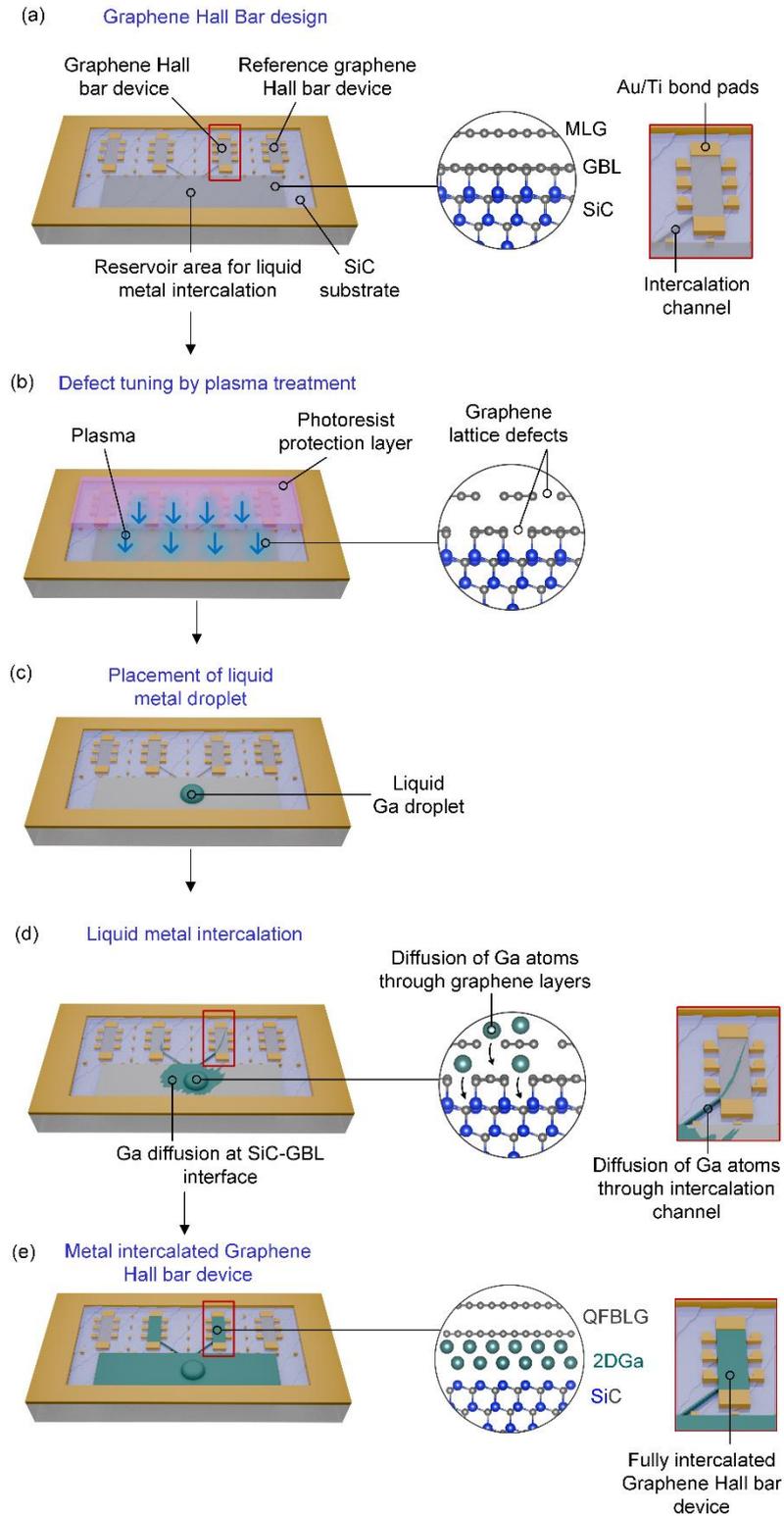

*Figure 2. Schematic design of liquid metal intercalation of epitaxial graphene Hall bars[11]. **a)** Graphene Hall bar design: Schematic representation of the graphene Hall bar device on a SiC substrate, including a reservoir area for liquid metal intercalation and reference Hall bar structures. **b)** Defect tuning by plasma treatment: controlled plasma treatment applied to selectively introduce lattice defects in the graphene layer. **c)** placement of liquid metal droplet: a liquid gallium (Ga) droplet is deposited onto the reservoir area, initiating the intercalation process. **d)** Liquid metal intercalation: Gallium atoms diffuse through the graphene layers and migrate along the SiC-graphene buffer layer (GBL) interface. **e)** Formation of the metal-intercalated graphene Hall bar: intercalated device consisting of decoupled quasi-freestanding bilayer graphene (QFBLG) and a confined gallium layer (2DGa) between the QFBLG and the SiC substrate.*

Following the defect engineering step, a droplet of liquid gallium is placed onto the reservoir area (Fig. 2c), where intercalation (Fig. 2d) occurs through the controlled application of pressure onto the metal droplet[12]. Gallium atoms diffuse through the graphene lattice defects (Fig. 2d), migrating towards the interface between the graphene buffer layer and the Si-face of the SiC substrate[12]. During this diffusion process, gallium atoms are arranged in a (1×1) periodicity with respect to Si(0001)[8], effectively transforming the epitaxial graphene into quasi-freestanding bilayer graphene (QFBLG)[8,12]. We assume that the gallium intercalation into epitaxial graphene is driven by interfacial energy minimization. The SiC(0001)/buffer layer interface is energetically unfavorable due to the high lattice strain in the (6√3 × 6√3) R30° reconstruction, partially $sp^3$-hybridized C-atoms [16] and unsaturated Si dangling bonds, which cause a high interfacial energy.

Upon intercalation, the graphene buffer layer transforms into graphene, relieving strain and restoring full $sp^2$ hybridization. Simultaneously, gallium saturates Si dangling bonds, further stabilizing the interface. This process reduces interfacial energy, facilitating metal confinement beneath graphene.

The gallium intercalation propagates from the reservoir area through the intercalation channels into the Hall bars. The final structure (Fig. 2e) represents a fully intercalated graphene Hall bar device. The intercalation dynamics strongly depend on the surface topography of the SiC substrate, a factor that will be discussed in detail later.

## Implementation and observation of metal intercalation in epitaxial graphene Hall bar devices

Figure 3a shows an optical microscope image of the lithographically fabricated graphene Hall bar structures designed for metal intercalation. The overall arrangement includes reference Hall bars and Hall bars specifically intended for metal intercalation, whereas the reservoir area is highlighted by a boxed region. Due to the high optical transparency of graphene, the Hall bars and intercalation channels cannot be clearly distinguished with standard optical microscopy[17].

The lithographically defined epitaxial graphene Hall bar devices are positioned such that the 4H-SiC terraces deviate by a few degrees from the [11$\bar{2}$0] crystallographic direction, with each device measuring approximately (400 × 1200) µm². The reference Hall bars are used for magneto-transport measurements, allowing a direct comparison with intercalated Hall bars to identify deviations in the quantum Hall effect (QHE) or the emergence of additional electronic phenomena. Note, that such contamination as dust particles appeared locally, but did not affect the intercalation process or magneto transport measurements itself.

Furthermore, the overall process has been carried out in a nitrogen atmosphere to prevent oxidation of the liquid metal. The real-time progression of the gallium intercalation at room temperature was monitored under an optical microscope, and video recordings of this process are provided in the supplementary materials (Fig. S3, V.1 , see SI).

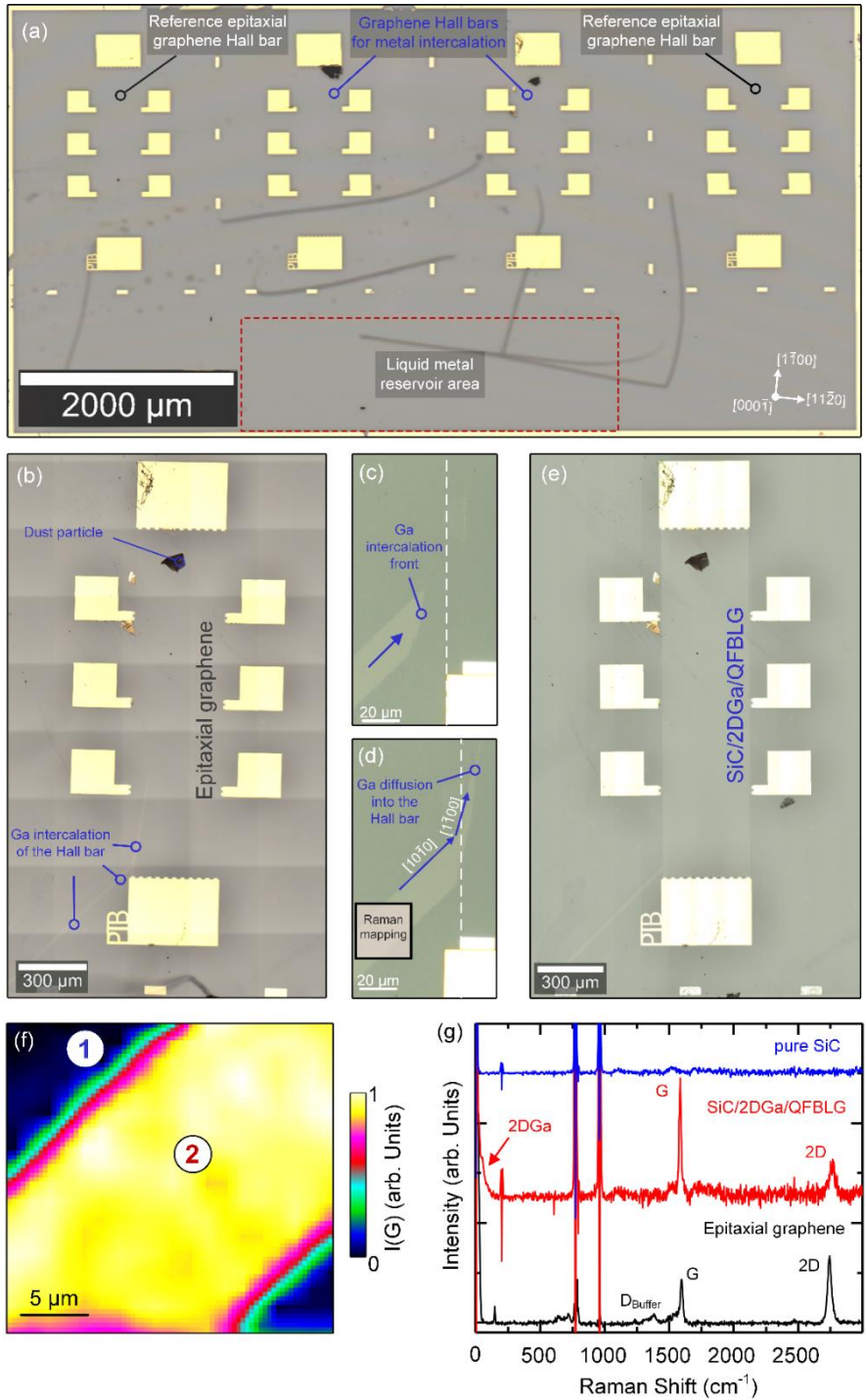

**Figure 3.** *a)* *Optical microscopy image of the fabricated Hall bar array, showing reference epitaxial graphene Hall bars and Hall bars designed for metal intercalation. The liquid metal reservoir area, where gallium is deposited for intercalation, is outlined in red. The crystallographic directions of the SiC substrate are indicated.* *b-d)* *Gallium intercalation observed at different time periods* *e)* *Fully intercalated Hall bar* *f)* *Raman intensity mapping of the G-band of intercalated QFBLG of the intercalation channel.* *g)* *Raman spectra of pure SiC away from the Hall bar device (blue), Ga-intercalated QFBLG (red), and epitaxial graphene (black) as reference.*

# Metal diffusion pathways and kinetics

Figures 3b–e show microscopy images capturing different stages of the intercalation process over time. The optical contrast of the epitaxial graphene and the intercalated graphene clearly reveals a strong contrast difference, which is indicative of gallium incorporation during and after the intercalation (Fi. 3b and d)[12]. Figure 3b already shows a contrast difference, indicating the gallium intercalation into the Hall bar through the intercalation channel. Figures 3c and d illustrate the final stages of intercalation, showing gallium approaching the Hall bar and diffusing into it. The optical microscopy image indicates that the intercalation started right after applying pressure onto the liquid droplet. Unlike other metals, gallium is already in its liquid state at room temperature, allowing intercalation to proceed even without additional heating. Here, we observed a successful intercalation of the reservoir region of approx. 15-20 min (see video V.2, SI). The intercalation dynamics were analyzed by tracking the progression of gallium diffusion over time. Based on the changes observed between Figures 3b and 3e, the complete intercalation of the Hall bar is estimated to take approximately one day, though a more precise determination requires further investigation. The migration of gallium atoms from the intercalation channel into the epitaxial Hall bar, spanning a ~120 μm distance between the intercalation fronts shown in Figures 3c and 3d, was observed to take approx. 7 minutes. Given that the total length of the intercalation channel is ~800 μm, we estimate that gallium diffusion through the channel takes approx. 47 min.

From these observations, we conclude that the intercalation speed depends on the alignment of the intercalation channel relative to the SiC terraces. As the channel orientation shifts, the dominance of the $[1\bar{1}00]$ plane diminishes, likely due to the increasing number of terrace steps that must be overcome, while the $[10\bar{1}0]$ plane serves as the primary diffusion pathway for the metal atoms. As a result, gallium diffusion between terraces is significantly hindered, due to increased potential barriers. Upon entering the Hall bar, the diffusion path of the gallium atoms abruptly changes, realigning along the $[1\bar{1}00]$ plane, suggesting a strong impact of the surface topography on the intercalation dynamics.

This directional preference for intercalation suggests that the fabrication process must account for terrace orientation to optimize the homogeneity of intercalation across the Hall bar device. The kinetics of metal diffusion is strongly affected by the surface potential of the SiC terraces. The observed diffusion of gallium at room temperature reveals an anisotropic behavior, predominantly occurring along the $[1\bar{1}00]$ direction, i.e., along the SiC terraces. This indicates diffusion along the terraces is preferred, guiding the movement of intercalated atoms at the interface between graphene and the SiC. Similar diffusion properties were also observed during the intercalation of lead (Pb) at 320°C beneath the graphene buffer layer on SiC[18], demonstrating that this phenomenon is not exclusive to gallium and epitaxial graphene.

The observed diffusion behavior aligns with expectations based on the Ehrlich–Schwoebel barrier[19], which hinders cross-terrace migration. Thus, we conclude that the observed anisotropic diffusion of gallium atoms is described by a two-dimensional diffusion-coefficient matrix, reflecting the effective confinement to the surface plane (i.e., the z-direction vanishes).

Specifically,

$$D = \begin{pmatrix} D_{xx} & D_{xy} \\ D_{yx} & D_{yy} \end{pmatrix} \qquad (1)$$

in which $D_{xx}$ and $D_{yy}$ represent the diffusion of gallium atoms along and orthogonal to the SiC terraces, respectively. The off-diagonal components $D_{xy}$ and $D_{yx}$ represent cross- or coupled diffusion processes between the two directions and are neglected for simplicity. Concretely,

$$D_{xx} = \frac{a^2}{z} v_\perp e^{\left(-\frac{\Delta E_\perp}{k_B T}\right)}, \; D_{yy} = \frac{a^2}{z} v_\| e^{\left(-\frac{\Delta E_\|}{k_B T}\right)} \qquad (2) \text{ and } (3)$$

where $a^2$ is the lattice spacing between neighboring surface sites, $z$ is the coordination number (i.e., the number of possible jump directions), $v_\|$ ($v_\perp$) is the attempt frequency for jumps along (perpendicular) the terrace, and $\Delta E_\|$ and $\Delta E_\perp$ the corresponding activation energy. The additional term $\Delta E_{ES}$ is the Ehrlich–Schwoebel barrier, which raises the effective activation energy $\Delta E_\perp$ for crossing terrace steps (see Eq. 4), resulting in $D_{yy} < D_{xx}$.

$$\Delta E_\perp = \Delta E_\| + \Delta E_{ES}, \qquad (4)$$

Hence, in our model, $\Delta E_\perp$ reflects the combined effect of the intrinsic diffusion barrier $\Delta E_\|$ relevant for in-plane jumps and the Ehrlich–Schwoebel barrier $\Delta E_{ES}$, which raises the potential energy required to cross terrace steps. Notably, if the Ehrlich–Schwoebel barrier becomes large (e.g., due to step bunching during graphene growth) then the diffusion matrix $D$ can become dominated by $D_{yy}$, reflecting strongly hindered edge-crossing diffusion. At higher intercalation temperatures, we assume that the dominance of the Ehrlich–Schwoebel barrier in $D_{yy}$, and thus the resulting anisotropic diffusion, may be significantly smaller.

The impact of step-edge barriers has been explored in prior studies, including DFT calculations on hydrogen as hetero-atom for adatom diffusion on a 3C-SiC(111) surface[20]. These studies reveal that on-terrace diffusion exhibits activation barriers of approximately 1.15–1.35 eV, while the step-edge crossing barrier (Ehrlich–Schwoebel barrier) is significantly higher, ranging from 1.65–1.75 eV. Based on these findings, we infer that similar energy differences could influence atomic diffusion during the intercalation process of graphene on SiC. To gain a deeper understanding of these dynamics, future studies should employ DFT calculations to systematically investigate confined gallium diffusion and quantify the relevant activation barriers.

Raman mapping was carried out to investigate the intercalation channels after gallium diffusion (Fig. 3d, black-colored box and Fig. 3f). Figure 3f presents the lateral intensity distribution of the G peak across the mapped area, clearly reflecting the shape of the intercalation channel. Raman spectra collected from two different regions (blue and red) are shown in Fig. 3g. The black-colored Raman spectrum in Figure 3g represents epitaxial graphene before gallium intercalation, measured on the reservoir area (measurement location not shown in Fig. 3).

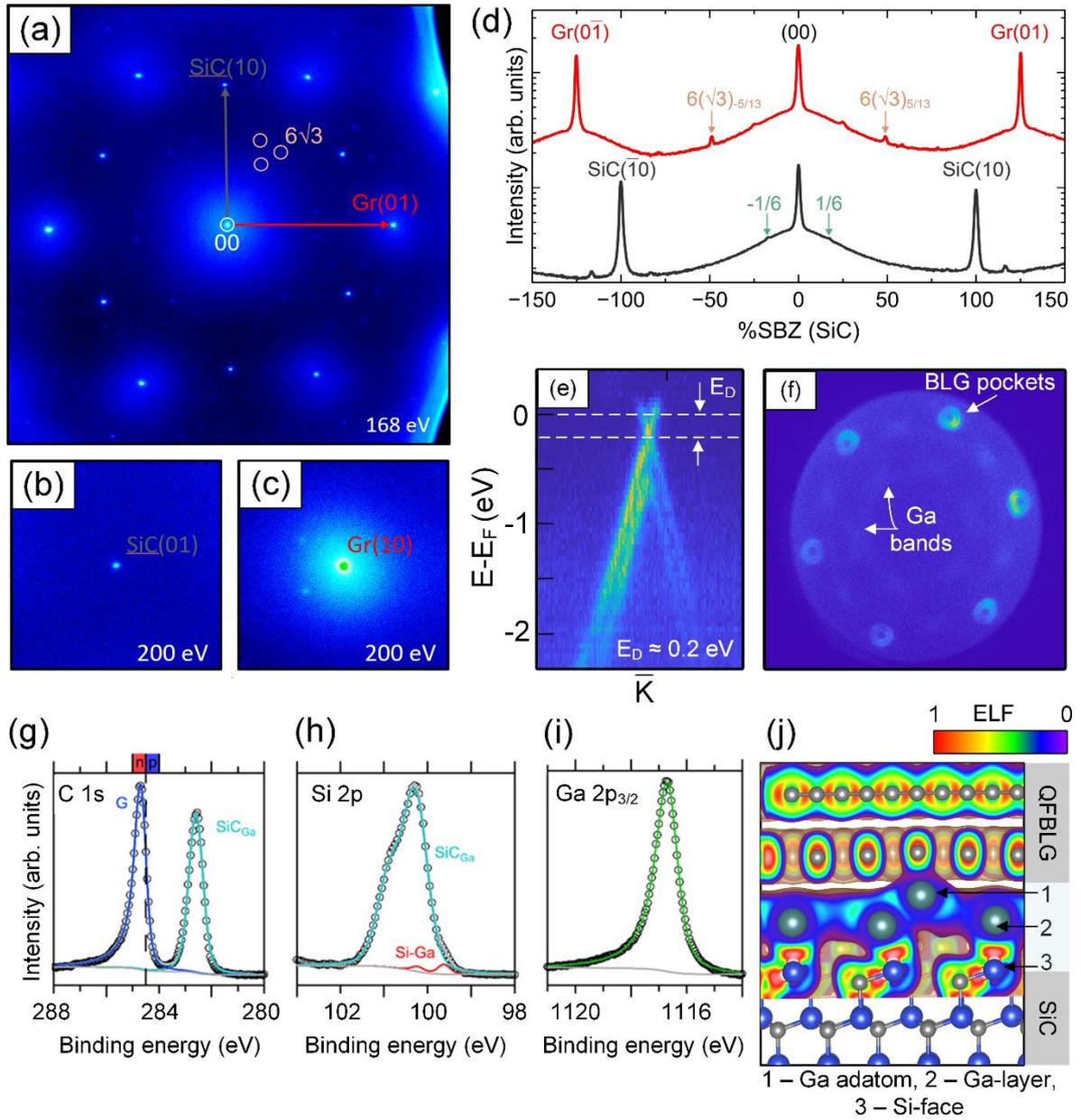

**Figure 4.** *a) SPA-LEED image of the Ga intercalated graphene on SiC acquired at 168 eV, showing the first-order SiC and Gr diffraction spots. b, c) Zoom in around the SiC and Gr spots at 200 eV. d) High-resolution spot profiles along the SiC (black) and Gr (red) direction. The curves are offset in y-axis for better visibility. e) Electronic band structure measured along the high-symmetry K direction. f) Fermi surface map showing the emergence of gallium-derived bands and bilayer graphene (BLG) pockets, confirming the impact of intercalation on the electronic structure. g) C 1s core level spectrum, of intercalated QFBLG h) Si 2p spectrum revealing Si-Ga binding energy i) Ga $2p_{3/2}$ spectrum, confirming the presence of gallium within the intercalated structure. j) DFT calculated Electron Localization Function (ELF). ELF mapping shows the electronic charge distribution in QFBLG with intercalated gallium. The first intercalated gallium layer exhibits strong localization near the Si-face, forming a covalently polarized bond, whereas in between gallium layer and the additional adatom exhibit more delocalized metallic behavior. The color scale indicates the degree of electron localization, where red represents highly localized electrons and blue indicates delocalized states.*

The blue-colored Raman spectrum represents the subtracted signal from a pure SiC reference spectrum, confirming the complete removal of epitaxial graphene after the etching process used for Hall bar lithography. Furthermore, the Raman mapping of the gallium intensity peak shows no detectable gallium signatures outside the intercalation channel, indicating that no deintercalation of gallium has occurred (Fig. S3, see SI). This observation underpins the stability of the intercalated gallium layer after the intercalation process. The Raman spectrum of epitaxial graphene before intercalation (Fig. 3g, black) exhibits characteristic G and 2D peaks associated with monolayer graphene, along with broad and flattened phonon bands from the buffer layer in the spectral range of 1200 cm$^{-1}$ to 1600 cm$^{-1}$ [21]. Following gallium intercalation, the graphene buffer layer background disappears, as indicated by the red spectrum in Fig. 3g, accompanied by an increase in intensity of the G peak at approx. 1586 cm$^{-1}$. Furthermore, the intensity and shape of the 2D peak undergo notable changes, reflecting the decoupling of the graphene buffer layer and the transition of epitaxial graphene into QFBLG [8,22]. The observed decrease in the 2D peak intensity is attributed to strong doping effects resulting from charge transfer between QFBLG and the intercalated gallium layers. The characteristic low-energy phonon modes of the gallium layer appear near the Rayleigh line at approx. 50 cm$^{-1}$ [23]. Importantly, no D peak around 1350 cm$^{-1}$ is observed, indicating the absence of lattice defects in QFBLG confirming the high quality of the LiMIT graphene.

## Spectroscopic characterization of Ga-intercalated QFBLG

The gallium intercalation of epitaxial graphene has been further validated through ARPES, SPA-LEED, and XPS measurements, complementing Raman spectroscopy. These measurements were conducted on large Ga-intercalated epitaxial graphene samples without structured Hall bars, where the sample quality is expected to be comparable.

Figure 4a shows a high-resolution SPA-LEED image recorded on the Ga-intercalated EG sample. The bright Gr spots indicate successful Ga intercalation. Notably, the Gr spots, as well as the (00) spot, are accompanied by a coherent background, as clearly seen in Figure 4(c) — the so-called bell-shaped component—which is a hallmark of free-standing graphene[24].

Moreover, the faint remnants of the 6×6 and 6√3 spots (also see Figure 4c and 4d) suggest the decoupling of the majority of the buffer layer, leading to the formation of QFBLG. However, the absence of Ga-induced superstructures indicates a 1×1 saturation of the interface (or Ga termination of the Si bonds), that has also been observed for other metal-intercalated graphene systems[25].

Electronic changes of QFBLG were investigated by ARPES measurements. Figure 4e presents the E–k dispersion at the K̄ point of QFBLG after Ga-intercalation around the K̄ point in the Brillouin zone, where the characteristic bilayer graphene π bands are visible. The observed splitting of the π bands originates from the interlayer interaction between the two graphene layers[26]. Notably, the downward shift of the Dirac point to $E_D \approx 200$ meV indicates charge carrier doping, which results from charge transfer between the confined 2D gallium layer and the two graphene layers above. The intercalated gallium layer at the interface serves as an electron donor, effectively introducing n-type doping into QFBLG.

The Fermi surface in Fig. 4f illustrates the first Brillouin zone of bilayer graphene, identified by the characteristic electron pockets at the high-symmetry K and K' points. Additionally, a weak background signal around Γ reveals the presence of gallium-derived bands. Briggs et al. assign these bands as near-free-electron-like states, which disperse upward towards the K and K' points of graphene[8]. However, these states do not directly hybridize with the Dirac states of graphene. Instead, the observed shift of the Dirac point confirms an electron transfer from the confined 2D gallium layer to the graphene layers, leading to n-type doping[8].

Figures 4g-i illustrate representative XP spectra of a different Ga-intercalated graphene sample. The C 1s signal is comprised of two components, corresponding to carbon atoms that are bound either in the quasi-freestanding graphene bilayer (G) or in the SiC bulk below the Ga layer (SiC$_{Ga}$). The position of the graphene component ($E_B$ = 284.6 eV) indicates slight n-doping (Fig. 4g), which is in agreement with the ARPES results. The binding energy of the bulk signal ($E_B$ = 282.6 eV) exhibits a notable shift of 1.1 eV to a lower binding energy compared to that observed for non-intercalated epitaxial graphene[27]. This can be attributed to an altered surface band bending, which occurs concurrently with the rearrangement of the surface bonding configuration. This phenomenon has been observed on numerous occasions in the context of metal intercalation of epitaxial graphene layers[18].

Furthermore, no signal from pristine MLG is detected in the Si 2p and C 1s spectra (Fig. 4g and 4h), indicating that the entire XPS measurement area is homogeneously intercalated.

A similar shift is evident in the bulk component of the Si 2p spectrum depicted in Figure 4h (SiC$_{Ga}$ at $E_B(Si\ 2p_{3/2})$ = 100.3 eV), which originates from silicon atoms bound within the bulk. Furthermore, an additional component, Si-Ga, was included in the analysis to reproduce the shoulder observed at low binding energy ($E_B$ = 99.6 eV), which can be assigned to the topmost Si atoms bound to the intercalated Ga layer. It is noteworthy that the surface component is considerably sharper than the corresponding bulk component. This phenomenon has been previously observed for Si-Pb bonds[18]. It is postulated that this is due to an enhanced screening of the core hole for bonding to the metallic Ga. Figure 4i depicts the Ga 2p$_{3/2}$ spectrum, which exhibits a single asymmetric component at 1116. 6 eV, in good agreement to the reported value for elemental Ga ($E_B$ = 1116.67 eV)[28]. The asymmetric line shape indicates that the 2D Ga layer retains its metallic character. We note that the absence of oxide signals underlines the protective effect of the graphene layers in the case of this sample.

A detailed analysis of the electron localization function (ELF) by using density functional theory (DFT) reveals distinct metallic characteristics among the intercalated gallium layer (Fig. 4j). The ELF mapping shows the electronic charge distribution in QFBLG with intercalated gallium layer and gallium adatom onto SiC surface [0001]. The first intercalated gallium layer exhibits strong localization near the Si-face, forming a covalently polarized bond, whereas in between gallium layer and the additional adatom exhibit more delocalized metallic behavior. The color scale indicates the degree of electron localization, where red represents highly localized electrons and blue indicates delocalized states.

# Magneto-transport measurements of Ga-intercalated QFBLG

Gallium intercalation is a highly dynamic process, where the lateral change of intercalated layers can differ significantly. This arises from the fact that mono-, bi-, and trilayers of gallium are energetically stable in confinement, while the growth process is primarily governed by the surface topography of the substrate [8]. SiC terrace edges play a crucial role in the growth process of confined metal layers during intercalation, leading to the formation of multilayer islands (up to 3 to 4 layers) that are energetically stabilized at step edges[8].

Figure 5a presents an SEM image of a phase boundary between epitaxial graphene and SiC/2DGa/QFBLG, which were measured on large Ga-intercalated epitaxial graphene samples. Here, the non-intercalated graphene monolayer exhibits a homogeneous dark contrast in the upper part (Fig. 5a), whereas the intercalated region shows pronounced contrast variations. Similar observations have been reported by El-Sherif et al., where the number of intercalated gallium layers was quantified through correlation analysis in SEM imaging[29]. While we do not apply this method here, the SEM contrast suggests strong dependence of the Ga interface layer from substate steps (mostly horizontal, marked by blue arrows) as well as graphene nanowrinkles (vertically connecting steps).

Local sheet resistance variations were investigated using four-point probe STM transport measurement in both linear and square tip configurations (Fig. 5b)[30]. No significant resistance changes were observed across epitaxial graphene (Positions 1–4) and at the phase boundary along SiC terraces, where resistance ranged between 0.9 and 2.3 kΩ/sq. However, at terrace edges (marked with blue arrows, Fig. 5a), the resistance increased significantly, reaching values

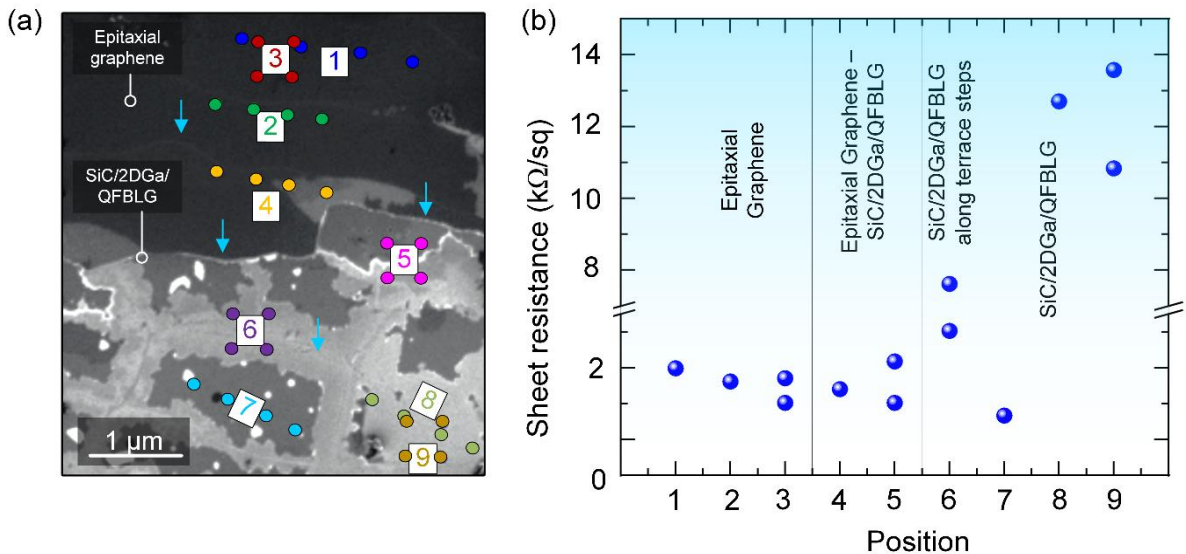

**Figure 5** *a) Scanning electron microscopy (SEM) image of the sample with color-coded measurement positions. Resistance measurements were performed in both linear and square configurations at different locations: epitaxial graphene (Positions 1–3), the transition region between epitaxial graphene and intercalated SiC/2DGa/QFBLG (Position 4), and Ga-intercalated graphene (Positions 5–9). The observed contrast variations in the intercalated regions may be attributed to differences in the number of intercalated gallium layers. Terrace steps on 4H-SiC are marked with blue arrows b) Position-dependent resistance extracted from four-point probe STM measurements*

between 2.3 and nearly 14 kΩ/sq (Positions 6, 8, and 9). Notably, the high pressure of the STM probe tips may locally modify the intercalated gallium layer or introduce defects, potentially affecting the measured resistance.

A comparable sheet resistance to epitaxial graphene was observed in Position 7 within SiC/2DGa/QFBLG on a terrace. Generally, these findings suggest that terrace edges can act as scattering centers for charge carriers in metal-intercalated layers, with possible contributions from variations in layer thickness or intrinsic lattice defects. A deeper understanding of the transport behavior related to local variations within the intercalated layer requires a more sophisticated investigation. However, at least one conductive graphene layer generally spans small interface defects, resulting in only minor variations in large-scale transport behavior. Intercalated areas further behind the phase boundary are likely to contain even less defects due to relaxation. Since no insulating areas where observed the transport behavior of the device (spanning several hundred μm) at room temperature will experience an average effect of interface variations.

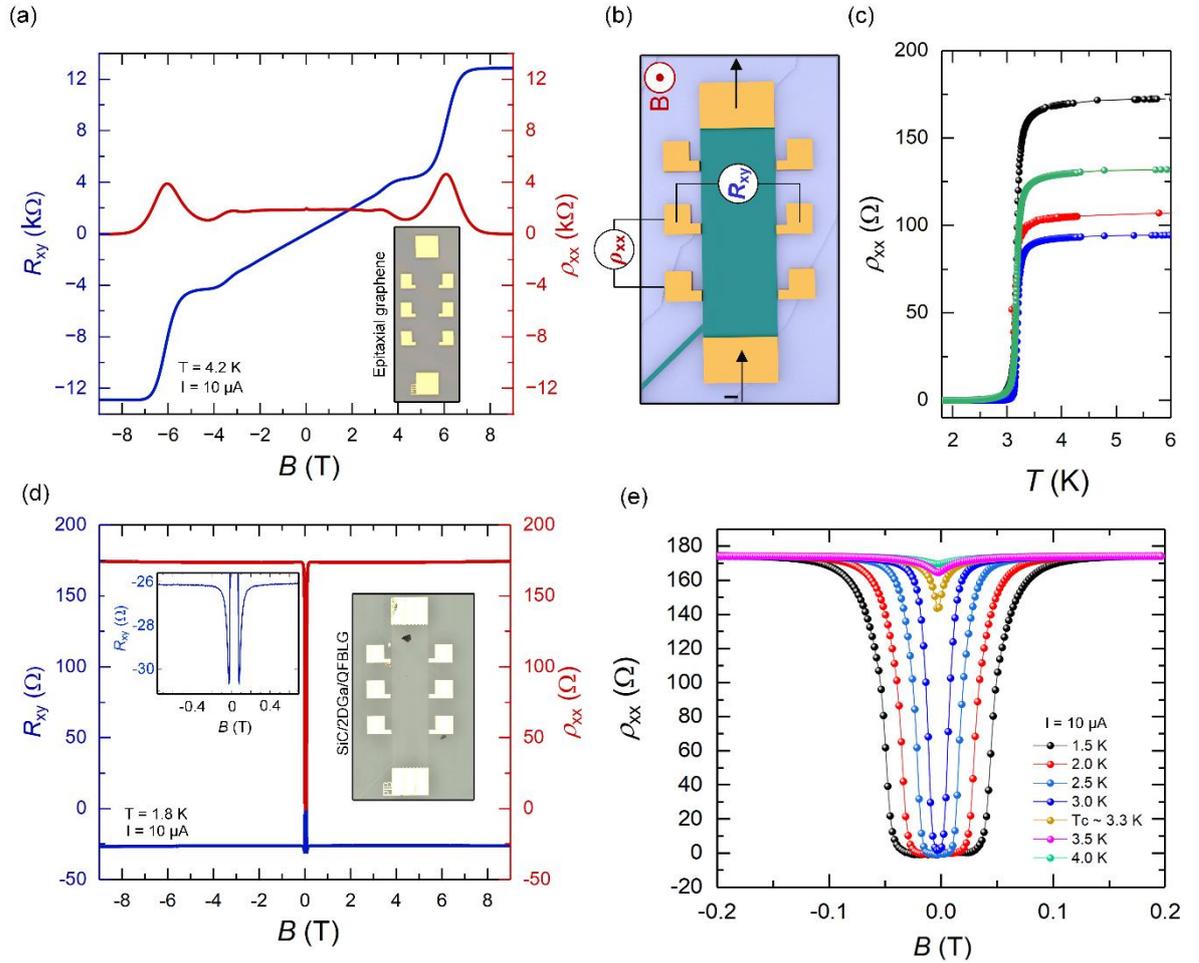

**Figure 6.** *a) Quantum Hall effect of epitaxial graphene at T = 4.2 K and I = 10 μA. Blue (red) ordinate depicts the Hall resistance (longitudinal resistivity) b) Schematic of the measurement setup for longitudinal ($\rho_{xx}$) and Hall ($R_{xy}$) resistances under an external magnetic field B. c) Temperature-dependent longitudinal resistance ($\rho_{xx}$) of Ga-intercalated quasi-freestanding bilayer graphene (SiC/2DGa/QFBLG). The superconducting phase in gallium of SiC/2DGa/QFBLG is observed at $T_{c,onset} \approx 3.5$ K. Blue (red) ordinate depicts the Hall resistance (longitudinal resistivity) d) Sweep of magnetic field strength in Ga-intercalated QFBLG at T=1.8 K and I=10 μA. e) Temperature-dependent sweep of magnetic field strength to investigate the change of $T_c$.*

The change of the electronic properties between epitaxial graphene and Ga-intercalated QFBLG was investigated by standard low-temperature magneto-transport measurements. Importantly, the intercalation channel was cut off using a diamond scriber to eliminate parasitic electronic effects from the metal reservoir region during the magneto-transport measurements in the Hall bar devices.

The Hall resistance curves $R_{xy}$ (blue curve) and the longitudinal resistivity $\rho_{xx}$ (red curve) of the devices are shown in Figure 6a for epitaxial graphene, and 6c, d, and e for the Ga-intercalated QFBLG Hall bar. Figure 6b shows schematically the setup for measuring $\rho_{xx}$ and $R_{xy}$. The transport measurements of the epitaxial graphene sample show all known typical QHE features (Fig. 6a). The fully developed quantum Hall resistance plateaus of $R_{xy} \approx 12.9$ kΩ related to the filling factor of $v = 2$ are observed at high magnetic fields ($B \geq 7$ T)[31]. In contrast, the longitudinal resistivity $\rho_{xx}$ vanishes in the same region.

Superconductivity in the large Ga-intercalated QFBLG Hall bar was confirmed through temperature-dependent measurements of the longitudinal resistivity $\rho_{xx}(T)$ using a current of $I = 10$ μA. Figure 6c presents four distinct $\rho_{xx}(T)$ measurements taken using four different contact pad pairs collinear with the current in the gallium intercalated Hall bar device. The observed superconducting onset temperature $T_c^{onset} \approx 3.5$ K is approximately 0.5 K lower than the values reported earlier by Briggs et al., while the zero-resistance transition temperature $T_c^0 \approx 3.25$ K is in good agreement[8]. A detailed description of the occurrence of superconductivity in the confined gallium layer can be found elsewhere[8].

Briggs et al. reported that $T_c^{onset}$ is affected by surface topography, where terrace steps formed by large step bunching lead to a slight suppression of $T_c^{onset}$ [8]. Additionally, variations in the slope of $R(T)$ were observed depending on the current direction, with distinct behaviors for measurements performed parallel and perpendicular to step edges in small-step and large-step samples. In our study, we cannot directly resolve the influence of terrace step orientation on $T_c^{onset}$ and the slope of $\rho_{xx}(T)$, as the Hall bar devices orientation deviates by a few degrees from the $[11\bar{2}0]$ crystallographic direction averaging out directional effects. Instead, we assume that the primary influence of surface topography arises from inhomogeneities in the gallium layer distribution rather than terrace step height, given that the latter remains within a narrow range of 0.50 to 0.75 nm (Fig. 5a).

Figure 6c reveals no difference in the superconducting onset temperature or the slope among the four measured $\rho_{xx}(T)$ curves. However, slight variations in the zero-resistance transition temperature are observed, ranging between 3.2 K and 3.0 K (Fig. 6c). Notably, as previously discussed, the longitudinal resistivity $\rho_{xx}$ above $T_c^{onset}$ exhibits values between approximately 90 Ω and 180 Ω. We attribute both observations to potential inhomogeneities in the distribution of intercalated gallium layers acting as intrinsic defects and thus affecting scattering spots for charge carriers. A detailed study is currently under way to better understand the effect of Ga layer inhomogeneities. The variation of $\rho_{xx}(T,H)$ with magnetic field strength shows that the superconducting temperature $T_c$ decreases, and the temperature width of the transition broadens with increasing magnetic field $B$. The critical magnetic field, $B_{c2} \approx 100$ mT, observed in the macroscopic Hall bar device (Figure 6e), is comparable to the 130 mT reported by Briggs et al. when using a four-point probe setup[8]. Thus, the corresponding coherence length is consistent with reported values [8]. Our Ga films display a critical current $I_c$ ~200 µA (Fig. S4, see SI), justifying the use of currents in the 10 µA range for the characterization and measurements.

A significant advantage of the Hall bar geometry is its ability to probe both the longitudinal and Hall resistance in an external magnetic field, enabling a detailed investigation of quantum phenomena such as the quantum Hall effect (QHE), the anomalous quantum Hall effect (AQHE), and proximity-induced effects[9,31,32,33,34].

Figure 6d shows a pronounced change in both the longitudinal ($\rho_{xx}$) and transverse ($R_{xy}$) resistances as a function of the applied magnetic field in the Ga-intercalated QFBLG system. Notably, the QHE is entirely quenched. As the magnetic field increases beyond $B_{c2} \approx 100$ mT, superconductivity is suppressed, leading to the emergence of electrical resistance. However, rather than showing QHE characteristics, the longitudinal normal state resistivity changes little with field to 9T. The transverse resistance shows both symmetric- and antisymmetric-in-field components. We attribute the symmetric-in-field component to non-uniform currents in the Ga film (equivalent to misaligned Hall contacts, which mixes $\rho_{xx}$ into $R_{xy}$), and an antisymmetric-in-field component characteristic of conventional Hall effect (See supplementary material S5).

From the antisymmetric component of $R_{xy}$ we estimate the carrier density in the range $n_e$ = 1.7 - 2.6 x $10^{16}$ cm$^{-2}$, which is several orders of magnitude higher than for as-grown epitaxial graphene (typically n ≈ 1 x $10^{13}$ cm$^{-2}$) [35]. This likely implies strong n-doping in the quasi-freestanding graphene with the primary current path located within the two-dimensional gallium layer.

Interestingly, the measured Hall resistance of the Ga-intercalated QFBLG does not follow a linear dependence on the applied magnetic field. Instead, it remains nearly constant above the superconducting transition, even upon field reversal, indicating an unconventional response that deviates from classical expectations. A notable anomaly is also observed in the abrupt emergence of sharp peaks in the Hall resistance just prior to the onset of superconductivity at fields below 100 mT (Fig.6c, left inset). Similar anomalous transverse resistance effects have been observed in various superconducting systems and may arise from distinct underlying mechanisms. Segal et al. demonstrated that inhomogeneous superconductors exhibit transverse voltage components due to variations in the longitudinal and Hall resistivity concerning temperature and magnetic field, leading to an even-in-field transverse response independent of vortex motion[36]. Furthermore, Sengupta et al. revealed that electronic inhomogeneities in superconducting thin films induce a highly non-uniform current distribution, giving rise to transverse resistance at macroscopic length scales, even in structurally uniform samples [37]. Additionally, Xu et al. reported an anomalous transverse resistance in the topological superconductor β-Bi$_2$Pd, which they attributed to broken inversion symmetry at the interface, potentially linked to topological surface states[38].

The observed behavior in the Ga-intercalated QFBLG system may result from a combination of these effects. Spatial inhomogeneities in the intercalated gallium layers could lead to current path distortions, while interface-induced symmetry breaking might contribute to unconventional transport signatures. Further investigations, including additional transport measurements and theoretical modeling, are required to deconvolute these contributions and elucidate the dominant mechanism.

Overall, our approach demonstrates that the proposed lithographic method for epitaxial graphene, combined with post-lithography intercalation via specifically designed intercalation channels, enables the scalable fabrication of well-defined intercalated Hall bar devices. These devices are suitable for systematic magneto-transport investigations and offer a platform for exploring the electronic properties of confined 2D metal layers.

## Conclusion

We have successfully developed a lithographically controlled intercalation approach based on liquid metal intercalation for fabricating metal-intercalated graphene Hall bar devices. This method enables precise control over intercalation dynamics while preserving the integrity of confined metal layers. By integrating lithographic structuring with post-lithography intercalation through dedicated diffusion channels, we achieved scalable and reproducible device fabrication, overcoming processing-induced deintercalation and ensuring structural stability. The introduction of engineered intercalation channels is crucial for directing the diffusion of gallium, allowing for localized and controlled metal incorporation into defect-free Hall bar devices.

Magneto-transport measurements confirm superconductivity in Ga-intercalated quasi-freestanding bilayer graphene, with a superconducting transition at $T_c^{onset} \approx 3.5$ K. The significantly higher charge carrier concentration in 2DGa, which is several orders of magnitude above that of graphene indicates that the superconducting path is primarily located within the 2D gallium layer. Furthermore, we observe the absence of a quantized Hall effect, while the transverse resistance exhibits both symmetric and antisymmetric field components. We attribute the symmetric-in-field component to non-uniform currents within the gallium film, whereas the antisymmetric component reflects the conventional Hall effect.

The ability to precisely modulate metal diffusion at the nanoscale establishes a robust platform for tunable intercalated graphene systems, offering new opportunities for proximity-induced superconductivity, quantum transport studies, and electronic applications. Given its scalability and compatibility with existing fabrication techniques, this method could be extended to metal-intercalated graphene-based electronic devices, paving the way for the integration of functional quantum materials into next-generation superconducting and nanoelectronic circuits.

# Acknowledgement

This work was supported by the Deutsche Forschungsgemeinschaft (DFG) project Pi385/3-1 within the FOR5242 research unit, and the DFG Germany's Excellence Strategy–EXC-2123 QuantumFrontiers – 390837967.